\documentclass[amsmath, amssymb,aps,PRApplied,preprint,superscriptaddress]{revtex4-2}

\usepackage{graphicx}
\usepackage{epstopdf}
\usepackage{subfig}
\usepackage{dcolumn}
\usepackage{bm}
\usepackage{gensymb} 

\begin{document}


\title{Self-biased magnetoelectric Ni/LiNbO\textsubscript{3}/Ni for body embedded electronic energy harvesters}


\author{Tianwen Huang*}
\affiliation{Sorbonne Université, CNRS, Laboratoire de Génie Electrique et Electronique de Paris, 75252, Paris, France\\
Université Paris-Saclay, CentraleSupélec,  CNRS, Laboratoire de Génie Electrique et Electronique de Paris, 91192, Gif-sur-Yvette, France}
\email{tianwen.huang@sorbonne-universite.fr}

\author{Loïc Becerra}
\affiliation{Sorbonne Université, CNRS, Institut des NanoSciences de Paris, INSP, UMR7588, F-75005 Paris, France}
\author{Aurélie Gensbittel}
\affiliation{Sorbonne Université, CNRS, Laboratoire de Génie Electrique et Electronique de Paris, 75252, Paris, France\\
Université Paris-Saclay, CentraleSupélec,  CNRS, Laboratoire de Génie Electrique et Electronique de Paris, 91192, Gif-sur-Yvette, France}
\author{Yunlin Zheng}
\affiliation{Sorbonne Université, CNRS, Institut des NanoSciences de Paris, INSP, UMR7588, F-75005 Paris, France}
\author{Hakeim Talleb}
\affiliation{Sorbonne Université, CNRS, Laboratoire de Génie Electrique et Electronique de Paris, 75252, Paris, France\\
Université Paris-Saclay, CentraleSupélec,  CNRS, Laboratoire de Génie Electrique et Electronique de Paris, 91192, Gif-sur-Yvette, France}
\author{Ulises Acevedo Salas}
\affiliation{Sorbonne Université, CNRS, Laboratoire de Génie Electrique et Electronique de Paris, 75252, Paris, France\\
Université Paris-Saclay, CentraleSupélec,  CNRS, Laboratoire de Génie Electrique et Electronique de Paris, 91192, Gif-sur-Yvette, France}
\author{Zhuoxiang Ren}
\affiliation{Sorbonne Université, CNRS, Laboratoire de Génie Electrique et Electronique de Paris, 75252, Paris, France\\
Université Paris-Saclay, CentraleSupélec,  CNRS, Laboratoire de Génie Electrique et Electronique de Paris, 91192, Gif-sur-Yvette, France}
\author{Massimiliano Marangolo}
\affiliation{Sorbonne Université, CNRS, Institut des NanoSciences de Paris, INSP, UMR7588, F-75005 Paris, France}



\date{\today}

\begin{abstract}

In this study, we present the fabrication and characterization of $\rm Ni/LiNbO_{3}/Ni$ trilayers using RF sputtering. These trilayers exhibit thick Ni layers (10 microns) and excellent adherence to the substrate, enabling high magnetoelectric coefficients. By engineering the magnetic anisotropy of Nickel through anisotropic thermal residual stress induced during fabrication, and by selecting a carefully chosen cut angle for the $\rm LiNbO_{3}$ substrate, we achieved a self-biased behavior. We demonstrate that these trilayers can power medical implant devices remotely using a small AC magnetic field excitation, thereby eliminating the need for a DC magnetic field and bulky magnetic field sources. The results highlight the potential of these trilayers for the wireless and non-invasive powering of medical implants. This work contributes to the advancement of magnetoelectric materials and their applications in healthcare technology.

\end{abstract}


\maketitle


\section{Introduction}
The important research effort on magnetoelectric (ME) composites opens perspectives in different engineering domains for designing sensors, transducers, filters, and other devices based on the change of $P$ (electric polarization) by an excitation magnetic field (direct ME effect) or by the change of $M$ (magnetization) through an excitation electric field (converse ME effect). Particularly exciting is the possibility to direct electric power obtained by ME energy harvesters from a magnetic field excitation source to embedded medical components \cite{paluszek2015magnetoelectric}. In this biomedical context, recent articles have shown the feasibility that an embedded sensor chip can be powered by a neighbouring ME resonator activated, through the human body, by a weak dynamic magnetic field working in a frequency range transparent through the human body (hundreds of kHz) while respecting the exposure limit value (1 Oe) \cite{malleron_experimental_2019,rupp_magnetoelectric_2019,zaeimbashi_ultra-compact_2021,chen_wireless_2022}.

Among the most promising ME devices, there are 2–2-type laminate composites constituted of ferroelectric (FE) compounds (ex.  $\rm BaTiO_{3}$, $\rm Pb(Zr, Ti)O_{3}$, $\rm Pb(Mg, Nb)O_{3}–PbTiO_{3}$) intimately connected to ferromagnetic (FM) and giant magnetostrictive materials (ex. $\rm CoFe_{2}O_{4}$, Terfenol-D, Galfenol) leading to an efficient cross-coupling via an intense mechanical elastic strain \cite{liang_magnetoelectric_2021}. The efficiency is measured by the ME coefficient, $\alpha_{E}$, expressed as $dE/dH = V_{ME}/(H_{ac} t_{p}$), where $t_{p}$ is the thickness of the piezoelectric substrate, $H_{ac}$ is an external small dynamic magnetic field ($\sim$ 1 Oe) under a static magnetic field $H_{DC}$ applied to the magnetostrictive layer, and $V_{ME}$ is the voltage induced across the electrodes of the piezoelectric transducer. It’s important to notice that $\alpha_{E}$ depends on the so-called piezomagnetic coefficient ($q= d\lambda/d H_{ac}$, where $\lambda$ is the magnetostriction strain) that is usually significant around the static magnetic bias field $\rm H_{bias}$ ($\sim$ several hundred Oe). Consequently, the ME devices need to be immersed in a static magnetic field generated by bulky permanent magnets or electromagnets, and excited by a small dynamic field generated by external coil or solenoid.

Although, in general terms, the requirements of a ME device depend on its functionality, a set of requirements is crucial for any application:
\begin{itemize}
    \item Reliability vs. cycling: fatigue and aging process of the adhesive binding is usually observed in (epoxy-)glued laminates undergoing fluctuating stresses and strains. Various epoxy-free methods have been developed in the last two decades to circumvent these limitations \cite{kumar_epoxy-free_2022}.
    \item Rare-earth-free and Lead-free structures to minimize environmental and health issues: PZT and NdFeB alloys should be avoided despite their optimal piezoelectric and magnetostrictive properties \cite{nan2008multiferroic}. Particular care must be taken for medical applications where materials function in vivo without eliciting detrimental responses in the body.
    \item 	Low-size and low-cost architectures: integrated permanent magnets to generate bias fields represent a serious bottleneck for the elaboration of compact and cheap devices. Self-biased systems permit ME effect in the absence of static magnetic bias $H_{DC}$ field \cite{zhou_self-biased_2016}. 
\end{itemize}

Recent articles reported promising results that fulfill all the aforementioned requirements. Nan et al. [8] elaborated compact, power efficient and self-biased magnetoelectric nano-electromechanical systems (NEMS) resonators constituted of $\rm AlN/(FeGaB/Al2O_3)  \times 10$. Self-biasing is due to magnetic domain wall motion that is not zero at zero bias magnetic field and changes the Young’s modulus. These sensors were produced by sputtering which guarantees reliable adhesion between the layers. Lage at al. adopted exchange bias between $\rm Mn_{70}Ir_{30}$ and $\rm Fe_{50}Co_{50}$ grown on piezoelectric AlN to self-bias ME cantilevers \cite{lage_exchange_2012}. Built-in stress during heterostructures preparation can induce self-biasing in 2-2 systems as shown in BTO-NCZF multilayers \cite{islam_magnetoelectric_2006} and in Ni/PZT/Ni multilayers \cite{poddubnaya2023magnetoelectric}, grown by co-firing and electrochemical deposition, respectively.

Self-biased ME effect can be obtained by tailoring the ferromagnetic phase or the interfacial coupling by playing with the FM composition, by designing a functional graded structure or by imposing a giant built-in stress along the constituent’s interface \cite{zhou_self-biased_2016}. Nevertheless, growth procedures are not easy-to-handle and self-biasing is obtained only at very low thickness and consequent low power.  

In this article, we report a new thick ME device 2–2-type composed of two $\rm \sim10 \ \mu m$ thick layers of magnetostrictive Nickel (Ni) and a piezoelectric $\rm \sim100 \ \mu m$ thick Lithium Niobate $\rm LiNbO_{3}$ (LNO). Two different commercially available $\rm LiNbO_{3}$ cuts were used and compared: $36\degree$ Y-cut and $163\degree$ Y-cut. This device is designed to fulfil the aforementioned requirements: Reliability vs. cycling is assured by the inter-layer physical binding obtained by magnetron sputtering growth. Self-biasing is ensured in a simple and reproducible way, i.e., by large in-plane residual stress arising from the thermal expansion coefficient mismatch of Ni and $\rm LiNbO_{3}$.

The dimensions of the samples are designed in order to get the resonant ME effect with work frequencies ranging between 160 kHz and 321 kHz, i.e., in the transparency range through the human body. These properties joined with the non-toxicity of the two compounds make the $\rm Ni/LiNbO_{3}/Ni$ system particularly well suited for biomedical-related applications \cite{zhou_self-biased_2016}. Admittedly, since nickel allergy is still a potential health problem affecting to 10 to 20 percent of the world’s population \cite{thyssen_metal_2010,jacob_nickel_2009}, encapsulation with a medical catheter can be envisaged.

\section{Experimental}

\subsection{Sample preparation}

\begin{figure}[htbp]
\centering
  \includegraphics[width=\linewidth]{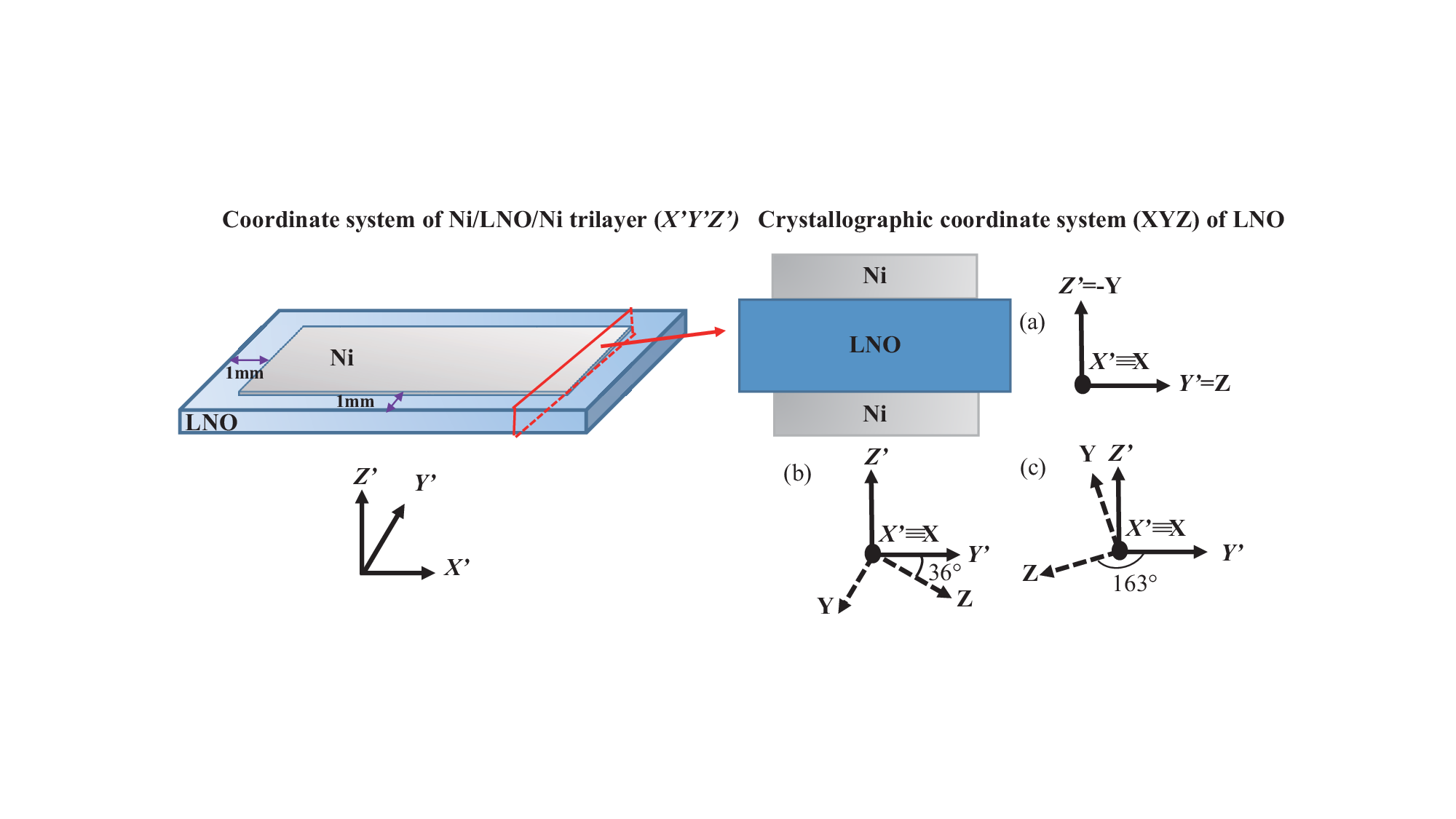}
  \caption{Illustration of different cuts of single crystal $\rm LiNbO_3$ substrate, (a) Y-cut, (b) 36\degree \ Y-cut, (c) 163\degree \ Y-cut.}
  \label{fig:Fig1}
\end{figure}

Fig.~\ref{fig:Fig1} shows the proposed ME laminated samples in which Ni polycrystalline films ($\rm \sim10 \ \mu m$) are deposed on a piezoelectric double side polished $\rm LiNbO_{3}$ layer (single crystal) in $36\degree$ or $163\degree$ Y-cuts of $tp=100 \rm \ \mu m$ thickness. Given its ferromagnetic behavior and electrical conduction, the Ni films have been deposed on each face by RF sputtering from a high purity Ni target ($\rm Ni > 99.95$$\%$), and act both as magnetostrictive layers and as electrodes for the $\rm LiNbO_{3}$.

The crystallographic X-Y-Z axes of the $36\degree$ and $163\degree$ Y-cuts $\rm LiNbO_{3}$ substrates  have been identified using in-plane X-ray diffraction (XRD) for detecting the ($2\overline{1}0$) peak intensity representing the X-axis ($\lambda\sim 1.540593 \ \rm \AA$). The observations have shown that for each $\rm LiNbO_{3}$ substrate, the crystallographic X-axis is parallel to the $\rm X^\prime$ (length) direction. Thus, according to the IEEE standard \cite{IEEE}, we could denote $36\degree$ and $163\degree$ Y-cuts used more precisely as $\rm (YXl) 36\degree$ and $\rm (YXl) 163\degree$ respectively. Before the sputtering deposition process, each $\rm LiNbO_{3}$ substrate has been chemically prepared through an acetone cleaner in an ultrasonic bath and the surfaces have been in situ cleaned with an argon plasma. The sputtering process on each $\rm LiNbO_{3}$ substrate face has been performed within a chamber pressure under $10^{-3}$ mbar of argon atmosphere at room temperature and with a supply power of 500 W. In this way, Ni films with a thickness of approximately 10 $\rm \mu m$ were deposited on both sides of a $\rm LiNbO_{3}$ substrate without continuous substrate rotation. The growth rate achieved was approximately 55.5 nm/min. Moreover the Ni target and the substrates are in a planar configuration during depositions. As shown in Figure 1, two gaps of 1 mm between the Ni films and the edges of the $\rm LiNbO_{3}$ substrate have been imposed to avoid a potential electrical connection between both Ni films.

\begin{figure}[htbp]
  \includegraphics[width=\linewidth]{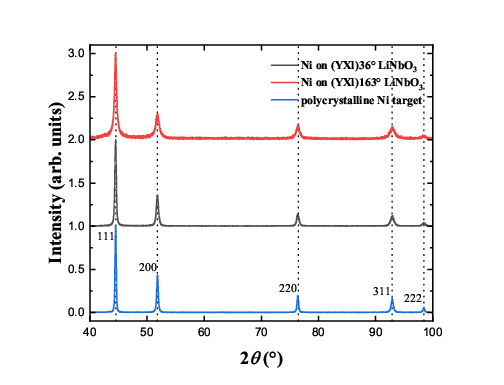}
  \caption{XRD patterns of the deposited Ni films.}
  \label{fig:Fig2}
\end{figure}

Fig.~\ref{fig:Fig2} shows the XRD patterns of a deposited nickel film compared with the pattern of the referential polycrystalline Ni target. It can be observed that each deposited Ni film on the $\rm LiNbO_{3}$ substrate behaves similarly to polycrystalline. Furthermore, the crystallite sizes of the Ni films on the $\rm (YXl) 163\degree LiNbO_{3}$ substrate are smaller than those on the $\rm (YXl) 36\degree LiNbO_{3}$ substrate.

\subsection{Experimental measurements and discussion}

\subsubsection{ME characterization}

\begin{figure}[htbp]
  \includegraphics[width=\linewidth]{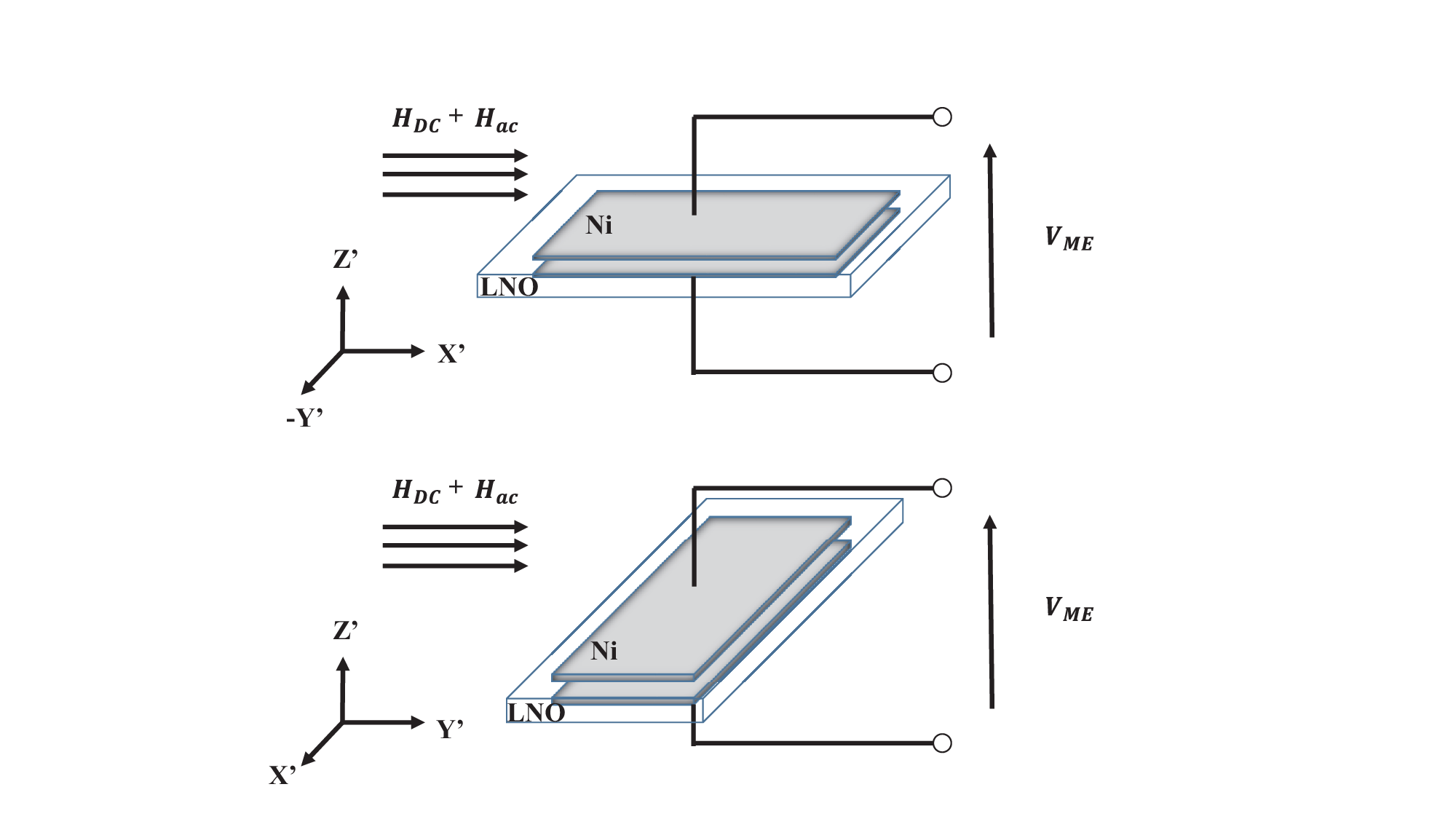}
  \caption{Schematic diagrams of $\rm Ni/ LiNbO_{3}/Ni$ trilayer function in LT mode.}
  \label{fig:Fig3}
\end{figure}

Fig.~\ref{fig:Fig3} depicts two magnetic excitation configurations used in LT mode for each ME $\rm Ni/LiNbO_{3}/Ni$ composite. Each composite was excited by an external magnetic field applied longitudinally (L-mode) in the $\rm X^\prime$ or $\rm Y^\prime$ in-plane direction, and the electric output voltage was polarized transversely (T mode) in the $\rm Z^\prime$ direction. 
As previous studies have shown \cite{zhou_tunable_2012,yang_resonant_2008}, the ME coefficient $\alpha_{E}$ is mainly influenced by the transverse piezoelectric coefficient $d_{3i}^{P}$. Consequently, the cuts of $\rm (YXl) 36\degree$ and $\rm (YXl) 163\degree$ presenting high and approximate values, i.e., $d_{31}^{P}$ and $d_{32}^{P}$ = 18 pC/N, have been selected to favor our study of ME effect with isotropic piezoelectric effects along the $\rm X^\prime$ and $\rm Y^\prime$ directions, respectively (see Fig.~\ref{fig:Fig3}).

ME coefficient characterization has been carried out using the experimental setup described in Ref \cite{malleron_experimental_2019}. A copper coil consisting of hundred turns was used to apply a small harmonic magnetic signal $H_{ac}$ = 1 Oe (in RMS). This field was driven by an AC current from an arbitrary waveform generator (Rohde \& Schwarz HM8150). Simultaneously, a static magnetic field $H_{DC}$ was superimposed in the same direction of the small harmonic magnetic signal $H_{ac}$ by adjusting the distance d between two opposite permanent magnets mounted on a moving rail. The value of $H_{DC}$ that was related to d has been previously calibrated using a Gaussmeter with its probe positioned within the center of the copper coil. Each ME sample was then inserted inside the coil to apply the superimposed magnetic field during the ME coefficient characterization. A specific support was used for aligning the length or the width direction of ME samples with the applied magnetic field. Thus, the ME voltage output ($V_{ME}$) generated between the Ni layers of a ME sample was then measured using a digital oscilloscope (Keysight MSO7054A). Finally, the ME coefficient $\alpha_{E} = V_{ME}/(H_{ac} t_{p})$, can be obtained in dynamic regime (or quasi-static regime).

\begin{figure}[htbp]
  \centering
  \subfloat[]
  {\includegraphics[width=0.5\textwidth]{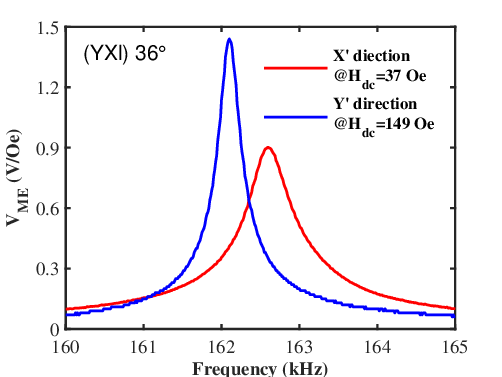}\label{fig:Fig4a}}
  \quad
  \subfloat[]
  {\includegraphics[width=0.5\textwidth]{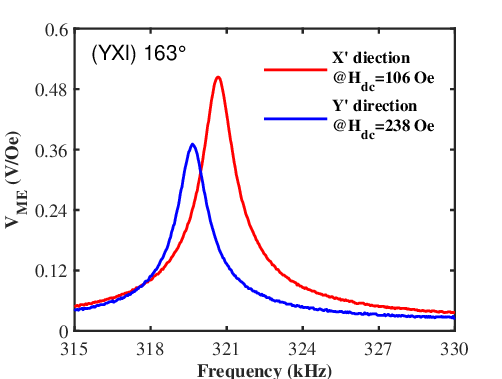}\label{fig:Fig4b}}
\caption{ME resonance measurement as a function of frequency for (a) $\rm Ni/(YXl)36\degree \ LiNbO_{3}/Ni$ and (b) $\rm Ni/(YXl)163\degree \ LiNbO_{3}/Ni$ samples in $\rm X^\prime$ and $\rm Y^\prime$ directions.}
\label{fig:Fig4}
\end{figure}

In order to characterize clearly the ME behaviors $\alpha_{E}$ versus $H_{DC}$ of the proposed $\rm Ni/LiNbO_{3}/Ni$ samples, a small harmonic magnetic signal $H_{ac}$ of 1 Oe was applied around the mechanical resonance (MR) frequency of each sample to obtain maximal mechanical oscillation in elastic phase. The first longitudinal MR was chosen and the calculated MR frequencies are 154.2 kHz and 348.1 kHz for the 20 mm- and 10 mm-wide samples, respectively \cite{wan_strong_2005}. As shown in Fig.~\ref{fig:Fig4}, the ME resonance frequency corresponding to the maximum of $V_{ME}$ was identified by ME voltage measurements performed by sweeping the frequency of the $H_{ac}$ signal, and we observed the measured ME resonance frequencies in two in-plane directions are close to the calculated MR frequencies as expected. After, the ME coefficient characterization was performed by sweeping $H_{DC}$ first from -710 to 710 Oe and from 710 Oe to -710 Oe. We changed the direction of $H_{DC}$ by permuting the permanent magnets in the position of $H_{DC} \approx  0 \:\rm Oe$. It is worth noting that this $H_{DC}$ sweep implied the first magnetization of the samples since they were prepared. 

In this study, we have investigated the impact of a $H_{DC}$ on the resonance frequency of the magnetoelectric (ME) response. Our findings reveal that setting the $H_{DC}$ to zero does not result in a significant alteration of the resonance frequency. Furthermore, we observed the shape of the ME coefficient curve remains unaffected by variations in the excitation frequency, despite the observed decrease in the ME response (see Fig.~\ref{fig:Fig4}).

\begin{figure*}[htbp]
  \includegraphics[width=\linewidth]{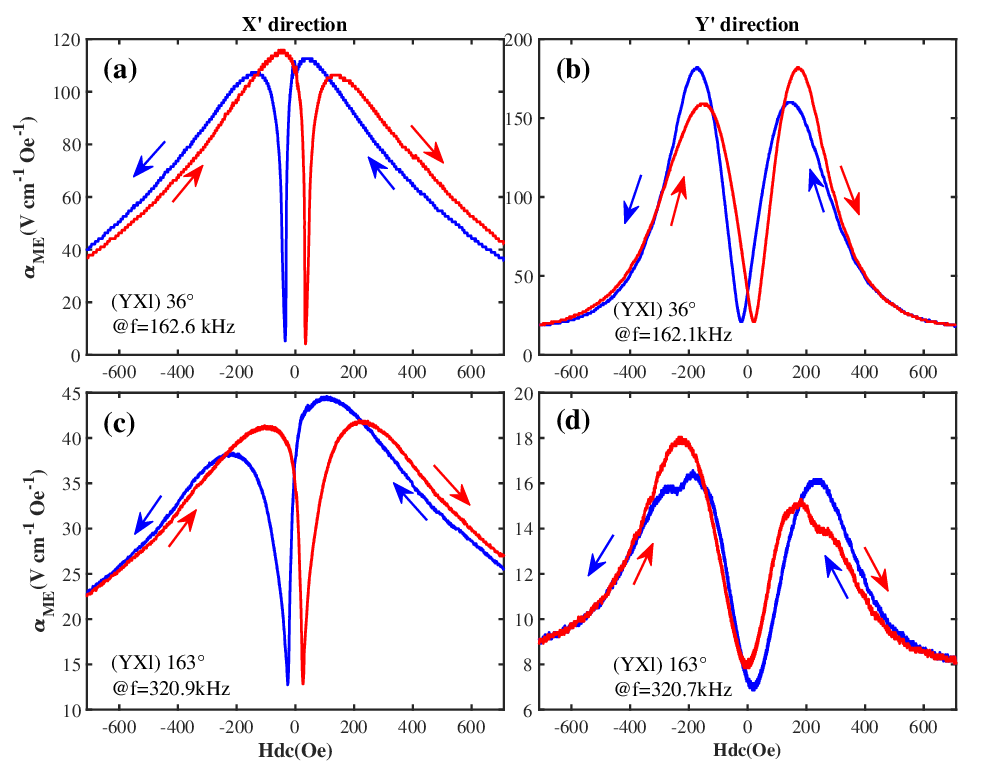}
  \caption{ME coefficient as a function of the $H_{DC}$ for $\rm Ni/(YXl)36\degree \ LiNbO_{3}/Ni$ and $\rm Ni/(YXl)163\degree \ LiNbO_{3}/Ni$ samples in $\rm X^\prime$ and $\rm Y^\prime$ directions. The measurements are performed at the ME resonance frequency of each sample.}
  \label{fig:Fig5}
\end{figure*}

As shown in Fig.~\ref{fig:Fig5}, each measurement has a maximum ME response at the magnetic bias field, $\rm H_{bias}$. We notice that $\rm H_{bias}$ values in the $\rm X^\prime$ direction are much lower than those in the $\rm Y^\prime$ direction: for the $\rm Ni/(YXl) 36\degree \ LiNbO_{3}/Ni$, $\rm H_{bias}$ is equal to $\pm$37 Oe and $\pm$149 Oe in the $\rm X^\prime$ and $\rm Y^\prime$ directions, respectively. For the $\rm Ni/(YXl)  163\degree \ LiNbO_{3}/Ni$, $\rm H_{bias}$ is equal to $\pm$106 Oe and $\pm$238 Oe in the $\rm X^\prime$ and $\rm Y^\prime$ directions, respectively. 

The main findings of our article are shown in Fig.~\ref{fig:Fig5} where “self-biased” behavior is reported for the applied magnetic excitation parallel to the $\rm X^\prime$ direction. Interestingly, at $H_{DC}$ = 0, $\alpha_{E}$ is at 94.20$\%$ of its maximum value for the sample grown on the $\rm (YXl) 36\degree \ LiNbO_{3}$ substrate and at 80$\%$ for the one grown on the $\rm (YXl) 163\degree \ LiNbO_{3}$ substrate. Moreover, the “self-biased” behavior is lost in the $\rm Y^\prime$ direction, where the $\alpha_{E}$ values are less than 40$\%$ of their maximum values.

Interestingly, by comparing their magnetoelectric responses with the ME trilayer composites Ni/PZT-5H/Ni in the literature \cite{bi_tunable_2011,pan_geometry_2009}, our results showed that our composite materials exhibited a significantly better magnetoelectric response at zero $H_{DC}$. These findings suggest that the composite material of $\rm Ni/LiNbO_{3}/Ni$ has the potential to be used in the development of advanced magnetoelectric devices.

\subsubsection{Magnetic anisotropies measurements: role played by the $LiNbO_{3}$ substrate}

In the following, we relate the ME response of our devices to the intrinsic magnetic properties of Ni films, and we show the important role played by the $\rm LiNbO_{3}$ substrate. For this purpose, we collected magnetic hysteresis loops by using a Vibrating Sample magnetometer (VSM). The magnetization ($M$) was measured at 300K as a function of $H_{DC}$ from 2 kOe down to 500 Oe (step = 10 Oe), then from 500 Oe to -500 Oe with high precision (step = 2 Oe), after from -500 Oe to -2 kOe, and finally a repetition in reverse. The samples were laser cut into $5\times 5 \ mm^{2}$ pieces to fit with the VSM sample holder. After measuring the magnetization of the trilayer sample, we isolated freestanding nickel films whose magnetization was measured by VSM.

\begin{figure}[htbp]
  \centering
  \subfloat[]
  {\includegraphics[width=0.5\textwidth]{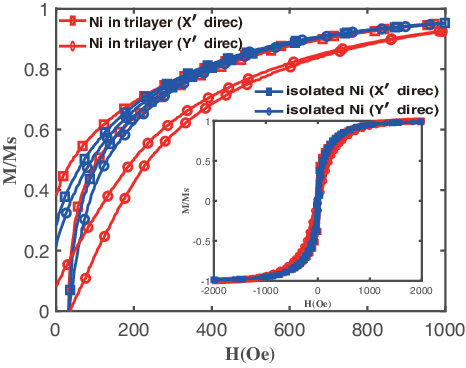}}\label{fig:Fig6a}
  \quad
  \subfloat[]
  {\includegraphics[width=0.5\textwidth]{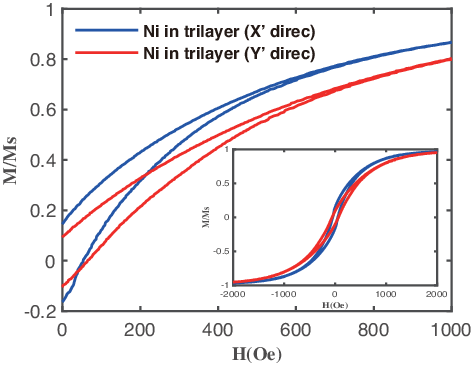}}\label{fig:Fig6b}
\caption{(a) Magnetization of nickel films in trilayer sample $\rm Ni/(YXl)36\degree \ LiNbO_{3}/Ni$ and its freestanding state in $\rm X^\prime$ and $\rm Y^\prime$ directions. (b) Magnetization of nickel films in trilayer sample $\rm Ni/(YXl)163\degree \ LiNbO_{3}/Ni$.}
\label{fig:Fig6}
\end{figure}

As shown in Fig.~\ref{fig:Fig6}, an anisotropic magnetization behavior is observed on trilayer samples with higher remanence and lower saturation field for the $\rm X^\prime$ direction as compared to the $\rm Y^\prime$ direction for the $\rm Ni/(YXl)36\degree \ LiNbO_{3}/Ni$ sample. A similar anisotropic magnetization behavior was also observed in the $\rm Ni/(YXl)163\degree \ LiNbO_{3}/Ni$ sample. It worth noting that the observed magnetic discrepancy between the two samples can be attributed to the difference in crystallite size as shown in Fig.~\ref{fig:Fig2}, affecting the magnetic coupling and alignment of magnetic moments within the films. Interestingly, the anisotropy of $\rm Ni/(YXl)36\degree \ LiNbO_{3}/Ni$ sample is strongly reduced in freestanding films indicating the important role played by residual stress imposed by the substrate. This allows us to infer that $\rm LiNbO_{3}$ substrates induce an anisotropic internal stress and a consequent magnetic anisotropy.

The magnetocrystalline anisotropy energy is estimated by computing the area between the $M–H$ curves along the $\rm X^\prime$ and $\rm Y^\prime$ directions. This yields 6.72 and 6.87 kJ/m\textsuperscript{3} for the $\rm Ni/(YXl)36\degree \ LiNbO_{3}/Ni$ and $\rm Ni/(YXl)163\degree \ LiNbO_{3}/Ni$ samples, respectively.

\subsubsection{Relation between magnetic and ME anisotropies}

In order to relate the magnetic anisotropy to the anisotropic ME response, we compare $M(H)$ cycles with the $\alpha_{E}$ field-dependence. Indeed, ME coupling combines piezomagnetic and piezoelectric effects. In this article, both effects are supposed to be linear. Following Ref \cite{zhou_tunable_2012} the ME coefficient can be written as the product of gradients of piezomagnetic and piezoelectric effects.
\begin{equation}
    \alpha_{E} = \left | \frac{\partial T}{\partial S} \times  \frac{\partial D}{\partial T} \times \frac{\partial E}{\partial D} \right | \times \frac{\partial S}{\partial H}
\end{equation}
Where $T$ is the stress, $D$ is the electric displacement field, $E$ is the electric field, $S$ is the strain.  Thus, the magnetic field dependence of the ME coefficient is such that $\alpha_{E} \propto \frac{\partial S}{\partial H} = \frac{\partial \lambda }{\partial H}$, where $\lambda$ is strain due to magnetostriction. Since $\lambda \propto M^{2}$, finally we obtained $\alpha_{E} \propto \frac{\partial M^{2}}{\partial H}$ as the relation connecting the ME response and the magnetization.

\begin{figure*}[htbp]
  \includegraphics[width=\linewidth]{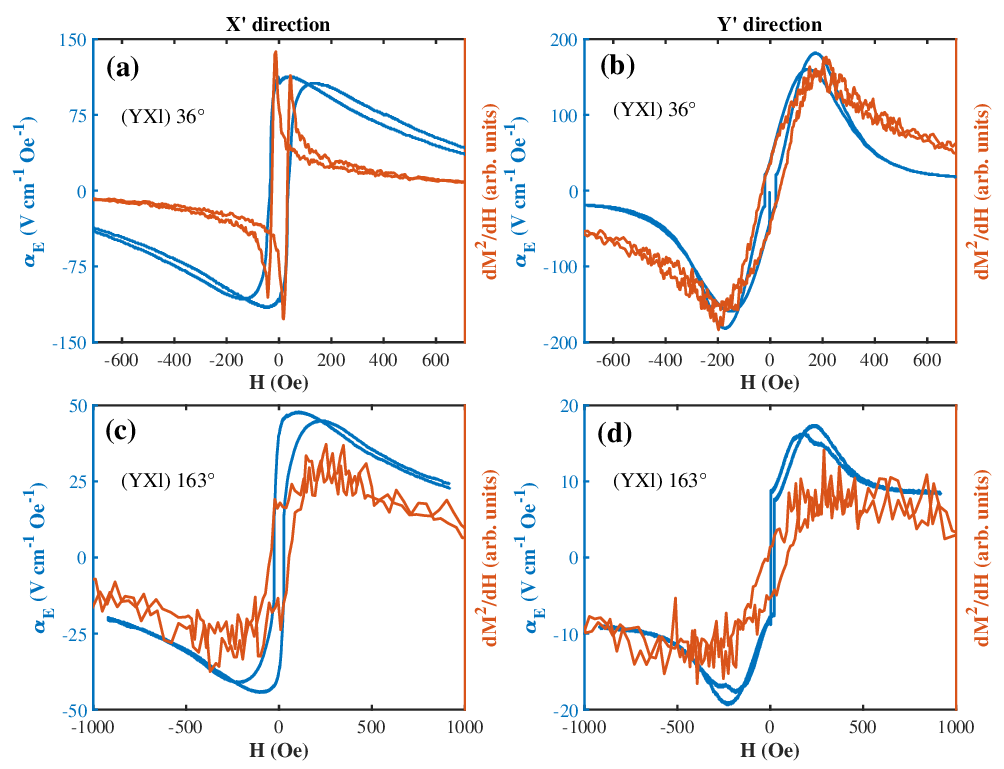}
  \caption{ME hysteresis loops and the $(dM^{2}/dH)-H$ curves of trilayer samples in $\rm X^\prime$ and $\rm Y^\prime$ directions.}
  \label{fig:Fig7}
\end{figure*}

As shown in Fig.~\ref{fig:Fig7}, we find a qualitative agreement between the $H$-dependence of ($\frac{\partial M^{2}}{\partial H}$) and the ME hysteresis loop (permutated the sign of $\alpha_{E}$ at the anti-resonances), attesting that the ME anisotropy behavior reflects the magnetic anisotropy of nickel films in contact with the $\rm LiNbO_{3}$ substrates. 

\subsubsection{Origin of magnetic and ME anisotropies}

Below, we show that the origin of the magnetic anisotropy resides in the mismatch between the thermal expansion coefficient (CTE) of Ni and $\rm LiNbO_{3}$ as the prevailing view \cite{huff_review_2022,truong_engineering_2022,mathews2023controlling}. Indeed, the samples are heated up during deposition due to exposure to the plasma and cooled down to ambient temperature ($\approx$ 300 K) after the sputtering growth procedure. Thus, the nickel films in trilayer samples are prone to residual stresses $\sigma_{r}$ related to thermal expansion effects, i.e., the very general formula.
\begin{equation}
    \sigma_{r,ii} = C_{e,iikl}(\alpha_{s}-\alpha_{f})_{kl} \bigtriangleup T
\end{equation}
 Where $\alpha_{s}$ and $\alpha_{f}$ are respectively the coefficients of thermal expansion of $\rm LiNbO_{3}$ substrate and Ni film, $\bigtriangleup T$ is the difference between ambient temperature and the temperature during the sputtering deposition process, and $C_{e,iikl}$ is a longitudinal coefficient in the elastic tensor of the Ni film. In the following, we focus on the in-plane biaxial (residual) stress induced in the Ni films, i.e. $\sigma_{r,xx}$  and $\sigma_{r,yy}$ . Since the CTE of $\rm LiNbO_{3}$ substrate in the $\rm X^\prime$ direction is higher than the CTE of Ni, as shown in Table~\ref{tab:table1} ($\alpha_{s}-\alpha_{f} > 0$), negative (compressive) residual stress grows up when cooling the sample. In contrast, the coefficient of thermal expansion of $\rm LiNbO_{3}$ substrate in the $\rm Y^\prime$ direction is less than the CTE of Ni, as shown in Table~\ref{tab:table1} ($\alpha_{s}-\alpha_{f} < 0$), leading to a positive (tensile) residual stress in cooling condition. Thus, the residual stress is anisotropic at room temperature leading to the observed magnetic anisotropy. Indeed, the order of magnitude of the residual stress can be evaluated by considering $\bigtriangleup T \sim 100$ K due to the growth procedure and the Young modulus of Nickel, E = 207 GPa. It turns out that the expected residual stress, $\sigma_{r}$, will be of the order of 50 MPa, with opposite signs in the $\rm X^\prime$ and $\rm Y^\prime$ directions. Hence, the expected magnetic anisotropy energy resulting from these two in-plane stresses is given by:
 
 \begin{equation}
 K_{u,\sigma } = 2\times \frac{3}{2} \times \sigma \times \lambda_{S} 
 \end{equation}
 
 with $\lambda_{S}= -34\times 10\textsuperscript{-6}$ \cite{mathews2023controlling}. Thus, we estimated  $K_{u,\sigma } \sim$ 5.1 J/m\textsuperscript{3}, in line with the qualitative magnetic anisotropy energies from Fig.~\ref{fig:Fig6}.

\begin{table*}[htbp]
    \caption{CTE of nickel $\alpha_{f}$ and of LNO substrates $\alpha_{s}$}
    \begin{ruledtabular}
        \begin{tabular}{@{}lll@{}}
            & $\alpha$ in $\rm{X'}$ direction [$\rm{K^{-1}}$] & $\alpha$ in $\rm{Y'}$ direction [$\rm{K^{-1}}$] \\
            \hline
            Nickel  & $13.4 \times 10^{-6}$ & $13.4 \times 10^{-6}$ \\
            $\rm{(YXl)36\degree \ LiNbO_{3}}$ & $15.4 \times 10^{-6}$ & $10.23 \times 10^{-6}$  \\
            $\rm{(YXl)163\degree \ LiNbO_{3}}$ & $15.4 \times 10^{-6}$ & $8.18 \times 10^{-6}$  \\
        \end{tabular}
    \end{ruledtabular}
    \label{tab:table1}
\end{table*}

Indeed, it is well known that such values can induce magnetic anisotropies in much thinner Ni films (35 nm) as demonstrated in \cite{finizio_magnetic_2014} when stress is applied statically by an electrostatic (PNM-PT) actuator and in \cite{wu_electrical_2011} where stress is applied by deforming a $\rm LiNbO_{3}$ substrate.

\subsubsection{Probing stress-induced magnetic anisotropy by magnetic and structural measurements}

In the following, we corroborate our hypothesis of residual thermal stress-induced magnetic anisotropy through two distinct experiments conducted on our two samples. However, the (311) crystal plane intensity measured by the in-plane XRD method in the Ni fims of the $\rm Ni/(YXl)163\degree \ LiNbO_{3}/Ni$ sample was weak, leading to noisy results. Consequently, we focused our analysis on the $\rm Ni/(YXl)36\degree \ LiNbO_{3}/Ni$ sample in the following, which provided clearer and reliable data.

\begin{itemize}
 \item $M(H)$ curves under different temperatures
 
    In Fig.~\ref{fig:Fig8}, we report the magnetization cycles measured by VSM in the $\rm X^\prime$ and $\rm Y^\prime$ directions, as a function of the temperature. It turns out that the magnetization cycles are strongly temperature dependent along the $\rm Y^\prime$ direction: magnetic remanence is higher at 400 K than at 200 K. In the $\rm X^\prime$ direction high remanence is found at all the probed temperatures. This indicates that residual stress is released by heating leading to a more isotropic magnetic behavior at 400 K. This behavior recalls the one observed in Figure 6 where magnetization cycles of freestanding films are strongly modified in the $\rm Y^\prime$ direction with respect to clamped trilayers.

\begin{figure}[htbp]
  \centering
  \subfloat[]
  {\includegraphics[width=0.5\textwidth]{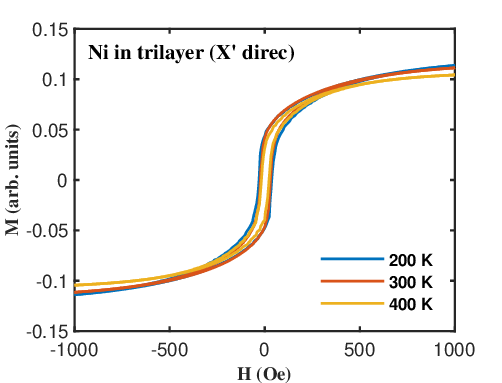}\label{fig:Fig8a}}
  \quad
  \subfloat[]
  {\includegraphics[width=0.5\textwidth]{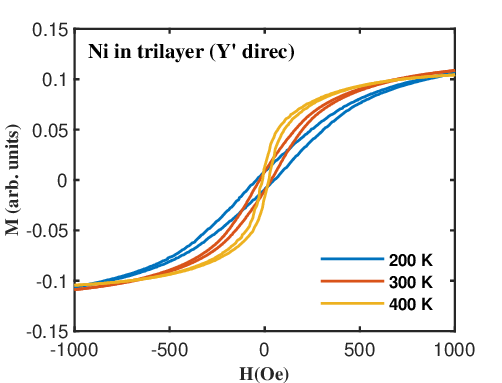}\label{fig:Fig8b}}
\caption{VSM Magnetization cycles were measured at different temperatures in the two in-plane directions of the $\rm Ni/(YXl)36\degree \ LiNbO3/Ni$ trilayer.}
\label{fig:Fig8}
\end{figure}

\item Determination of residual stresses using X-ray diffraction method

Finally, according to \cite{huff_review_2022}, we confirmed the existence of residual stresses by a direct and quantitative determination of atomic spacing changes of nickel film by in-plane XRD phi-scan along the $\rm X^\prime$ and $\rm Y^\prime$ directions, we compared the atomic spacing corresponding to the peak (311), $d_{311}$ , in the Ni film deposited on the $\rm LiNbO_{3}$ substrate with the freestanding Ni film after removal from the substrate. As shown in Fig.~\ref{fig:Fig9}, by a fit of the measured peaks (311) using the Pseudo-Voigt function, we could estimate $d_{311}$ in the $\rm X^\prime$ and $\rm Y^\prime$ directions. As shown in Table~\ref{tab:table2}, the measured negative (positive) atomic spacing change signifies the presence of a negative (positive) residual stress in the $\rm X^\prime$ ($\rm Y^\prime$) direction. Using the Young's modulus of 207 GPa and the Poisson's ratio of 0.29 for Ni material, a temperature change $\bigtriangleup T$ of approximately -70 K can be derived. However, the temperature change of -70 K involved is slightly lower than the estimated temperature change of -100 K from deposition. The difference could be attributed to the fact that grazing incidence in-plane XRD measurements primarily capture the top part of the Ni films, rather than the portion in contact with the $\rm LiNbO_{3}$ substrate.

\begin{figure}[htbp]
  \includegraphics[width=\linewidth]{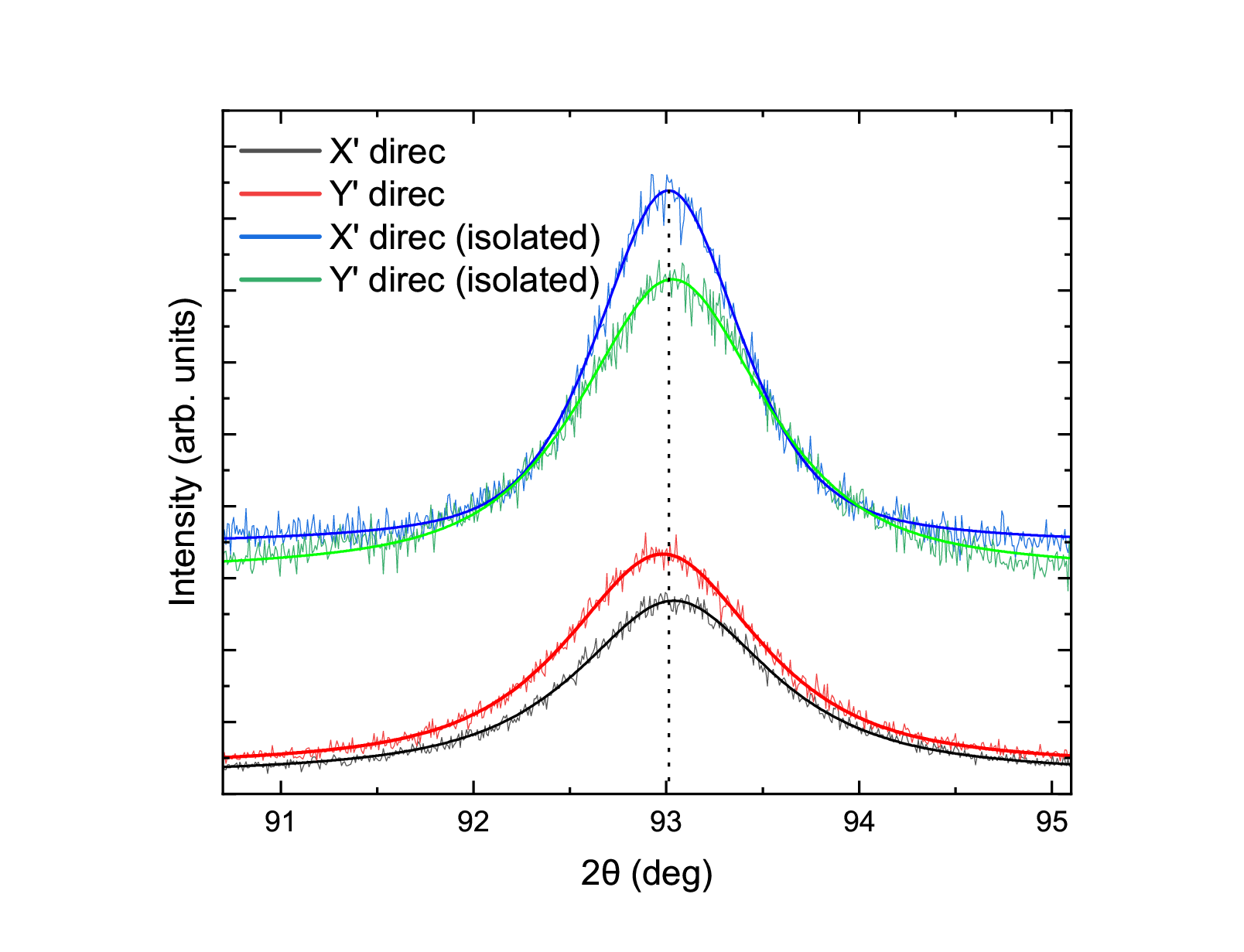}
  \caption{In-plane XRD patterns of the (311) peak of Ni from the sample comprised of $\rm (YXl)36\degree \ LiNbO_{3}$.}
  \label{fig:Fig9}
\end{figure}

\begin{table*}[htbp]
    \caption{In-plane strain determined by XRD of Ni (311) reflection along $\rm{X^\prime}$ and $\rm{Y^\prime}$ directions in Ni in trilayer and in isolated Ni layer. The standard error on the last significant digit of the values is indicated in parentheses.}
    \begin{ruledtabular}
        \begin{tabular}{ccccc}
            & \begin{tabular}[c]{@{}c@{}}Ni in trilayer\\ (X' direction)\end{tabular} & \begin{tabular}[c]{@{}c@{}}Isolated Ni\\ (X' direction)\end{tabular} & \begin{tabular}[c]{@{}c@{}}Ni in trilayer\\ (Y' direction)\end{tabular} & \begin{tabular}[c]{@{}c@{}}Isolated Ni\\ (Y' direction)\end{tabular} \\ 
            \hline
            $2\theta$ [degree] & 93.040(3) & 93.013(3) & 92.983(3) & 93.032(4) \\
            $d_{311}$ [$\mathrm{\AA}$] & 1.06246(3) & 1.06269(3) & 1.06296(3) & 1.06253(3) \\
            \begin{tabular}[c]{@{}c@{}}strain [\%]\end{tabular} & \multicolumn{2}{c}{-0.022(5)} & \multicolumn{2}{c}{0.041(6)} \\ 
        \end{tabular}
    \end{ruledtabular}
    \label{tab:table2}
\end{table*}

\end{itemize}

\section{Conclusion}

In this work, we have shown that $\rm Ni/LiNbO_{3}/Ni$ trilayers permit to envisage efficient self-biased magnetoelectric devices. The growth by magnetron sputtering of these lead-free and rare-earth-free trilayers guarantees a tight and resistant physical binding between the magnetostrictive and piezoelectric layers with respect to the epoxy-assisted adhesive binding that is prone to fatigue and aging process. Ni layers are thick (~10 $\rm \mu m$ each) inducing significant longitudinal deformations of the $\rm LiNbO_{3}$ layer ($\sim 100 \rm \ \mu m$) and consequent high ME output ($\sim$ 100 V/cm Oe) in the $\sim$ 100-300 kHz regime. By a judicious choice of the crystalline orientation of the $\rm LiNbO_{3}$ substrate (i.e., $36\degree$ Y-cut and $163\degree$ Y-cut), an anisotropic residual stress is induced spontaneously in the Ni layers during cooling after growth, as attested by X-ray diffraction and magnetometry measurements. We demonstrate that residual stresses are due to the thermal expansion coefficient mismatch of Ni and $\rm LiNbO_{3}$ that, in turn, induces a significant magnetic anisotropy and consequent self-biasing of the device. 

Thus, $\rm Ni/LiNbO_{3}/Ni$ has the potential to enhance its self-biased properties by controlling thermally the residual stress. These findings suggest that the $\rm Ni/LiNbO_{3}/Ni$ composite may have potential applications in the development of self-biased magnetoelectric devices, and also we observed the ME coefficient responses of $\rm Ni/LiNbO_{3}/Ni$ are better than the Ni/PZT-5H/Ni in the literature.

This study opens the way to a transformative leap forward in magnetoelectric technology for implanted medical devices. Indeed, self-biased $\rm Ni/LiNbO_{3}/Ni$ trilayer composites grown by physical tools could be the building block of a biocompatible, corrosion resistant, durable vs. cycling power supplies that could be charged by an external tiny ($\sim$ 1 Oe) oscillating magnetic field without the need of any DC field, avoiding bulky and expensive magnetic field sources. 

\begin{acknowledgments}
This research has received funding from the French National Research Agency under the project Biomen (Projet-ANR-18-CE19-0001). The authors acknowledge the staff of the MPBT (physical properties - low temperature) platform of Sorbonne Université for their support, and also the use of the laser cut with Femtosecond laser micromachining at the Institut de Minéralogie de Physique des Matériaux et de Cosmochimie (IMPMC), Paris. Part of this work was carried out thanks to the technological capabilities of the Salles Blanches Paris Centre (SBPC) network.
\end{acknowledgments}

\bibliographystyle{apsrev4-2}
\bibliography{references}

\end{document}